\documentclass{rrparticle}
\usepackage{graphicx}

\newcommand{\miktex}{\hbox{Mik\kern-.15em\TeX}}

\title{Classicalization and quantization of tachyon-like matter on (non)archimedean spaces}
\author[1,a]{Dragoljub D. Dimitrijevic}
\author[1]{Goran S. Djordjevic}
\author[1]{Milan Milosevic}
\affil[1]{Department of Physics, Faculty of Science and Mathematics, University of Nis,
\\Visegradska 33, P.O.Box 224, 18000 Nis, Serbia\\Email:$^a$ {\em ddrag@pmf.ni.ac.rs}}

\begin{document}
\maketitle
\begin{abstract}

We consider a class of tachyon-like potentials, inspired by string
theory, D-brane dynamics and cosmology in the context of classical
and quantum mechanics. Motivated by the trans-Plankcian problem in
the very early stage of cosmological evolution of the Universe, we
consider the theoretical role of DBI-type tachyon scalar field,
defined over the field of real as well as $p$-adic numbers, i.e.
archemedean and nonarchimedean spaces. To simplify the equation of
motion for the scalar field, canonical transformations are defined
and engaged. The corresponding quantum propagators in the Feynman
path integral approach on real and nonarchimedean spaces are
calculated and discussed, as are possibilities for a quantum
adelic generalization and its application.

\end{abstract}

\section{Introduction}

According to the theory of cosmological inflation, the inhomogeneities in our Universe have a quantum-mechanical
origin. This scenario is phenomenologically very appealing as it solves the problems of the standard hot big
bang model and in a natural way explains the spectrum of cosmological perturbations \cite{martin12}.

In addition, one of the main goals of quantum cosmology is to find
a quantum (gravitational) state which contributes to a consistent
description of inflationary period of the Universe evolution. When
accomplished, we could make predictions which should be more
consistent with observational data.

According to the inflationary scenario, the Hubble radius was (almost)
constant during inflation, the wavelength of a mode of
astrophysical interest was much smaller than the Hubble
scale at the beginning of inflation. Moreover, the modes are
initially sub-Planckian, not only sub-Hubble, their wavelength is
smaller than the Planck length. In this case the physics we are
familiar with is not applicable any more \cite{martin04, martin01}. That
leads us to idea to consider physical models on noncommutative and
non-Archemedean spaces \cite{we-vol, GLjDR, liu}.

The general conclusion of various investigations in this direction of
research is the observation that in the sub-Planckian (or
trans-Planckian) regime the precise description of the evolution
of the fluctuations is unclear. Various proposals are introduced
in order to better understand the picture in the trans-Planckian regime, such
as modification of dispersion relation by the trans-Planckian
effects \cite{martin01}, modification of uncertainty relation
\cite{kempf01}, or the introduction of a new fundamental scale
\cite{danielsson}.

In keeping with this direction, we consider the DBI-type tachyon scalar
field theory \cite{sen02} in cosmological context as an effective
field theory (taken from the string theory) which describes
rolling tachyons \cite{gibbons}. This effective field theory has
already been studied in the context of tachyonic inflationary
cosmology, where, e.g. one can get kinematically driven inflation,
the so-called "k-inflation" \cite{picon, garriga}. Although it is
argued that tachyonic inflation (on real numbers-spaces!) can not
be responsible for the last $60$ e-folds of inflation, it might be
possible that tachyonic inflation is responsible for an earlier
stage of inflation, at and "around" the Planck scale, which may be important
for the resolution of homogeneity, flatness and isotropy problems
\cite{kofman}. For interesting ideas of $p$-adic inflation and
$p$-adic rolling tachyons in a string theory context, see
\cite{Cline} and \cite{MoellerZwiebach}.

Here we will discuss the idea of using $p$-adic numbers ($p$ stands for a prime number) in investigation of
trans-Planckian effects in cosmology. Use of $p$-adic analy\-sis is motivated and based on the simple fact that results of
experimental and observational measurements always give some rational numbers. So, the field $R=Q_{\infty}$ of real numbers
and the field $Q_p$ of $p$-adic numbers are naturally put on the same footing, both are built by completion of the field
 of rational numbers with respect to the corresponding norms \cite{ostrowski}. $p$-Adic norm is a non-Archimedean (ultrametric) one \cite{vvz}.

$p$-Adic numbers in cosmology are usually connected with quantum
cosmology. Since quantum cosmology is used to describe the
evolution of the universe at a very early stage, i.e. it is
related to the Planck scale phenomena, then it is reasonable to
consider various geometries, in particular the non-Archimedean
one, which is closely connected to $p$-adic numbers and their
application \cite{we-vol, dragovich09}.

$p$-Adic quantum mechanics has been developed in two different
ways. Here we will use the one where the wave function is a
complex valued function of the $p$-adic variable, at least for an adelic approach (simultaneous treatment of real and
$p$-adic numbers) to quantum physics. This formulation is the
preferred one \cite{idaqp} and has a natural, quantum,
interpretation.

In order to explore theoretical models describing the early stage of
cosmological evolution of the Universe (e.g. period of inflation)
it is sufficient to start with spatially homogenous scalar fields
\cite{kofman,copeland,frolov,steer,fdp,ddmm}. Cosmological
perturbations are then introduced for the spatially homogenous
scalar field and spatially homogenous background metric. As has already been noted, we will
consider the DBI-type tachyon scalar field, following Sen's proposals and conjectures \cite{sen02},
concentrating on the equation of motion for the field. Because this equation is not so
simple even in the case of a spatially homogenous scalar field in
the flat space-time background, we will try to deal with it
imposing a canonical transformation treatment. This ``transition``
from field theory to a classical mechanical model \cite{fdp, kar} we will call
here ``classicalization``. Although it is an approach different from the idea recently proposed by Dvali and others \cite{dvali}, there are
interesting similarities in equation of motions of DBI Lagrangians
\cite{tetradis}.

We will show that for certain choices of generating function
of canonical transformation the equation of motion simplify
significantly. At this point one can involve Lagrangian of the
standard type which corresponds to the same equation of motion
and which is locally equivalent \cite{aip14, facta14}.
It allows us to quantize the simplified (toy) model, obtain quantum propagator and vacuum states
in $p$-adic and adelic framework, as well as discuss their basic physical consequences.

This paper proceeds as follows. In Section \ref{tach_cosmology} we
will briefly review tachyonic cosmology, after which Section
\ref{ct} deals with canonical transformation. In Section
\ref{choice} certain choices of generating function of canonical
transformation are introduced and presented through two examples.
Section \ref{generalization} deals with general remarks on
classically equivalent Lagrangians (``classicalization``), while
Section \ref{p-adic} presents quantization of the classical
models, i.e. their real and $p$-adic consideration and a short review
on a natural adelic generalization. We conclude in Section
\ref{conclusion} by discussing obtained constraints on possible
values of tachyon fields, time and constants appearing in two
tachyonic models we consider.

\section{Tachyonic Cosmology and Inflation}
\label{tach_cosmology}

The Lagrangian we are dealing with - the DBI-type Lagrangian - is of non-standard
type, it contains potential as a multiplicative factor and a term
with derivatives ("kinetic" term) inside the square root
\cite{sen02,garousi}
\begin{equation}
\label{} \mathcal{L}_{tach}={\mathcal L}(T,\partial_{\mu
}T)=-V(T)\sqrt{1+(\partial T)^2},
\end{equation}
\noindent where $T$ is tachyonic scalar field, $V(T)$ - tachyonic
potential, $(\partial T)^2 = g_{\mu\nu}\partial^{\mu }T\partial^{\nu }T$ and
$g_{\mu\nu}$ - components of the metric tensor, with "mostly positive signature".
The tachyon potential $V(T)$ has a positive maximum at $T=0$ and a minimum at $T_0$ with $V(T_0)=0$
($T_0$ can be either finite or infinite). General expression of the energy-momentum tensor
\begin{equation}
\label{EMT} T_{\mu \nu}={\mathcal L}_{tach}g_{\mu \nu}+
V(T)\frac{\partial_{\mu }T\partial_{\nu }T}{\sqrt{1+(\partial T)^2}},
\end{equation}
\noindent written in the form of the energy-momentum tensor for the ideal fluid defines pressure and energy density of
a fluid described by the tachyonic scalar field
\begin{equation}
\label{density} \rho = \frac{V(T)}{\sqrt{1+(\partial T)^2}},
\end{equation}
\begin{equation}
\label{pressure} p = {\mathcal L}_{tach}.
\end{equation}
\noindent The state parameter $w$ is then
\begin{equation}
\label{w} w = \frac{p}{\rho} = -1 - (\partial T)^2.
\end{equation}
\noindent One can define the so called effective sound speed $c_s$ (for the cosmological perturbations \cite{garriga})
\begin{equation}
\label{Cs} c_s^2 = \frac{\partial{\mathcal L}_{tach}}{\partial(\partial T)^2}
\left(\frac{\partial\rho}{\partial(\partial T)^2}\right)^{-1},
\end{equation}
\noindent which takes into account "friction" effects and it is a generic feature of theories with nonstandard Lagrangians
of this type.

Equation of motion in curved spacetime is \cite{frolov, gibbons}
\begin{equation}
\label{}
D_{\mu}\partial^{\mu }T-\frac{D_{\mu}\partial^{\nu }T}{1-(\partial T)^2}
\partial_{\mu }T\partial_{\nu
}T - \frac{1}{V(T)}\frac{dV}{dT} = 0,
\end{equation}
\noindent where $D_{\mu}$ is covariant derivative with respect to $g_{\mu\nu}$.
For the spatially homogenous tachyon field in flat
spacetime background the last equation is reduced to
\begin{equation}
\label{tachyonEOM} \ddot T(t) - \frac{1}{V(T)}\frac{dV}{dT} \dot
T^2(t) = -\frac{1}{V(T)}\frac{dV}{dT}.
\end{equation}
\noindent In this case one can
starts with the Lagrangian
\begin{equation}
\label{tachyonL}\mathcal{L}_{tach}(T,\dot{T}) =
-V(T)\sqrt{1-\dot{T}^2},
\end{equation}
\noindent from which equation of motion (\ref{tachyonEOM}) is
obtained. In the classical mechanical limit $T$ can be replased by $x$ 
and considered as a position variable in one-dimensional space \cite{fdp, kar}.
The equation contains a term proportional to $\dot T^2$,
which is, in general, less convenient to deal with. Note that even in the case of Minkowski spacetime background,
the equation of motion for tachyonic field contains term with $\dot T^2$,
which is not the case for theories with standard-type Lagrangians. Objects $w$ and $c_s$ in this case
correspond to each other in a simple manner \cite{bilic}
\begin{equation}
\label{w1} w = -1 + \dot{T}^2 = -c_s^2.
\end{equation}
%
%\begin{equation}
%\label{Cs1} c_s^2 = \frac{\partial{\mathcal L}_{tach}}{\partial\dot{T}^2}
%\left(\frac{\partial\rho}{\partial\dot{T}^2}\right)^{-1},
%\end{equation}
%
%\noindent and they correspond to each other in a simple manner
%
%\begin{equation}
%\label{} c_s^2 = -w.
%\end{equation}

\noindent Again, this kind of correspondence between $w$ and $c_s$ is not the case for the standard-type Lagrangian theories,
where the effective sound speed is constant and equal to one (in units where speed of light $c=1$) \cite{bilic}.

%\section{Friction Term}
%
%We are writing again equation of motion (\ref{tachyonEOM} for the later purpose
%
%\begin{equation}
%\label{tachyonEOM1} \ddot T(t) - \frac{1}{V(T)}\frac{dV}{dT} \dot
%T^2(t) = -\frac{1}{V(T)}\frac{dV}{dT}.
%\end{equation}
%
%If the friction term linearly depends on $\dot T$, the particle finally stops after a finite time

\section{Classical Canonical Transformation}
\label{ct}

In the light of the above discussion on tahyonic dynamics on (non)-Archeme\-dean spaces,
a natural question arises: is it possible to simplify the
equation of motion (\ref{tachyonEOM}) and, if it is possible, in which cases could that be done
\cite{fdp, ddmm, aip14, facta14}? Further, is it possible to quantize the models,
at least for some particular potentials? A new possible way is to try with
canonical transformation(s), which will be introduced and explained in this Section.

A classical canonical transformation is a change of the phase space variables
$(T, P)$ to a new $(\tilde{T}, \tilde{P})$, which preserves the Poisson bracket
\begin{equation}
\{T,P\}_{P.B.} = 1 = \{\tilde{T},\tilde{P}\}_{P.B.}.
\end{equation}
\noindent We will seek for unitary transformations of coordinate (field) $T$ and conjugate momenta $P$ at the classical level
\begin{equation}
\label{} T, P \quad \mapsto \quad \tilde{T}, \tilde{P},
\end{equation}
\begin{equation}
\label{} \mathcal{H}_{tach}(T,P) \quad \mapsto \quad \tilde{\mathcal{H}}_{tach}(\tilde{T},\tilde{P}),
\end{equation}
\noindent which also preserves form of Hamilton's equations
\begin{equation}
\label{}\dot{T} = \frac{\partial\mathcal{H}_{tach}(T,P)}{\partial P}
\quad \rightarrow \quad
\dot{\tilde{T}} = \frac{\partial\tilde{\mathcal{H}}_{tach}(\tilde{T},\tilde{P})}{\partial \tilde{P}},
\end{equation}
\begin{equation}
\label{}
\dot{P} = -\frac{\partial\mathcal{H}_{tach}(T,P)}{\partial T}
\quad \rightarrow \quad
\dot{\tilde{P}} = -\frac{\partial\tilde{\mathcal{H}}_{tach}(\tilde{T},\tilde{P})}{\partial \tilde{T}}.
\end{equation}
\noindent Note the conjugate momenta and (conserved) Hamiltionian are \cite{gibbons}:
\begin{equation}
\label{tachyonP}
P = \frac{\partial L_{tach}}{\partial \dot{T}} = \frac{\dot{T}}{\sqrt{1-\dot{T}^2}}V(T),
\end{equation}
\begin{equation}
\label{tachyonH}\mathcal{H}_{tach}(T,P) = \sqrt{P^2 + V^2(T)}.
\end{equation}
\noindent We will use a particular model for generating function $G$, which specifies point canonical transformation.
It is constructed as a function of new field $\tilde{T}$ and old momenta $P$
\begin{equation}
\label{can_trans}
G(\tilde{T}, P) = -PF(\tilde{T}),
\end{equation}
\noindent where $F(\tilde{T})$ is an arbitrary function of a new field. By definition,
the old coordinate $T$ and the new momenta $\tilde{P}$ are uniquely defined
\begin{equation}
\label{ct1}
T = -\frac{\partial G}{\partial P} = F(\tilde{T}),
\end{equation}
\begin{equation}
\label{ct2}
\tilde{P} = -\frac{\partial G}{\partial \tilde{T}} = P\frac{dF(\tilde{T})}{d\tilde{T}}.
\end{equation}
\noindent At the same time the new coordinate and old momenta can be expressed as
\begin{equation}
\label{ct3}
\tilde{T} = F^{-1}(T),
\end{equation}
\begin{equation}
\label{ct4}
P = \frac{1}{F^{\prime}(\tilde{T})}\tilde{P},
\end{equation}
\noindent where $F^{-1}(T)$ is an inverse function of $F(\tilde{T})$ and
$F^{\prime}$ denotes derivative with respect to $\tilde{T}$.
It is easy to check that Jacobian of this canonical transformation is equal to one and
the Poisson brackets are unchanged, as expected.
%
%\begin{equation}
%\label{}
%\{T,P\}_{P.B.} = \{\tilde{T},\tilde{P}\}_{P.B.} = 1.
%\end{equation}
%
%\noindent

Hamilton's equations become
\begin{equation}
\label{}
\dot{\tilde{T}} = \frac{1}{F^{\prime}}\frac{\tilde{P}}{\sqrt{\tilde{P}^2 + {F^{\prime}}^2V^2}},
\end{equation}
\begin{equation}
\label{}
\dot{\tilde{P}} = -\frac{1}{{F^{\prime}}^2}\frac{1}{\sqrt{\tilde{P}^2 +
{F^{\prime}}^2V^2}}[(F^{\prime})^2V\frac{dV(F)}{dF} -
F^{\prime\prime}P^2],
\end{equation}
\noindent while the equation of motion transforms to
\begin{equation}
\label{eom1}
\ddot{\tilde{T}} + \left(\frac{F^{\prime\prime}}{F^{\prime}} - F^{\prime}\frac{d\ln V(F)}{dF}\right)\dot{\tilde{T}}^2 +
\frac{1}{F^{\prime}}\frac{d\ln V(F)}{dF} = 0.
\end{equation}
\noindent Note that equation (\ref{eom1}) still contains term quadratic with respect to time derivative of a new
coordinate (field) $\dot{\tilde{T}}$ (like (\ref{tachyonEOM})). Let us stress that this procedure,
at the classical level, is formally invariant with respect to the choice
of the background number fields $R$ or $Q_p$.

\section{Choice of $F(\tilde{T})$}
\label{choice}

Up to now function $F(\tilde{T})$ was an arbitrary one. If the function $1/V(T)$ is integrable, function $F(\tilde{T})$
can be defined in such a way that its inverse function $F^{-1}(T)$ is equal to
\begin{equation}
\label{F}
F^{-1}(T) = \int_{}^{T}\frac{dT}{V(T)},
\end{equation}
\noindent where the lower limit of the integral can be chosen arbitrary. This particular choice enables us to simplify equation of motion (\ref{eom1}) significantly. In particular,
the second term in (\ref{eom1}) vanishes in this case, because the expression in parenthesis is identically equal to zero
\begin{equation}
\label{}
\frac{F^{\prime\prime}}{F^{\prime}} - F^{\prime}\frac{d\ln V(F)}{dF} = 0.
\end{equation}
\noindent So, using (\ref{F}), the equation of motion (\ref{eom1}) is reduced to the form
\begin{equation}
\label{eom2}
\ddot{\tilde{T}} + \frac{1}{F^{\prime}}\frac{d\ln V(F)}{dF} = 0.
\end{equation}
\noindent Note that equation (\ref{eom2}) now stands for the system without term
quadratic with respect to $\dot{\tilde{T}}$ (unlike (\ref{tachyonEOM}) and (\ref{eom1})).
The term disappeared after we imposed suitable chosen canonical transformation, i.e. choice (\ref{F}).

It is very important to emphasize that equation of motion (\ref{eom1}) in some cases can be obtained from the standard
type Lagrangians \cite{fdp, ddmm, musielak}. Moreover, equation (\ref{eom2}) in some cases can be obtained from the standard type Lagrangians
which are quadratic with respect to $\tilde{T}$ and $\dot{\tilde{T}}$, a situation very suitable for the Feynman path integral approach for quantization.
From several interesting tachyonic or tachyon-like potentials we have considered, we choose two of the most used class of potentials to
study here in more detail: $V(T)=e^{-\alpha T}$ and $V(T)=1/\cosh(\beta T)$ \cite{steer, garousi05}.

\subsection{Example 1: Exponential potential}

The case of exponential potential of a tachyonic field
\begin{equation}
\label{exp-pot}
V(T) = e^{-\alpha T}, \quad \alpha > 0 - const,
\end{equation}
\noindent is well known and motivated by string theory \cite{sen02}. The function $F^{-1}(T)$ becomes
\begin{equation}
\label{}
F^{-1}(T) = \frac{1}{\alpha}e^{\alpha T},
\end{equation}
\noindent which means that function $F(\tilde{T})$, using (\ref{ct1}), becomes
\begin{equation}
\label{}
F(\tilde{T}) = \frac{1}{\alpha}\ln(\alpha \tilde{T}).
\end{equation}
\noindent The full generating function (\ref{can_trans})
\begin{equation}
\label{}
G(\tilde{T}, P) = -PF(\tilde{T}) = -\frac{P}{\alpha}\ln(\alpha \tilde{T}),
\end{equation}
\noindent reduces equation of motion to the well known form
\begin{equation}
\label{}
\ddot{\tilde{T}} - \alpha^2\tilde{T} = 0.
\end{equation}
\noindent This equation of motion can be delivered from a quadratic Lagrangian
\begin{equation}
\label{quad1L}\mathcal{L}_{quad}(\tilde{T},\dot{\tilde{T}}) =
\frac{1}{2}\dot{\tilde{T}}^2 + \frac{1}{2}\alpha^2\tilde{T}^2.
\end{equation}

\subsection{Example 2: Potential $V(T)=1/\cosh(\beta T)$}

As in the previous case, with a similar motivation, we consider the potential
\begin{equation}
\label{cosh-pot}
V(T)=\frac{1}{\cosh(\beta T)}, \quad \beta > 0 - const,
\end{equation}
\noindent and the function $F^{-1}(T)$ becomes
\begin{equation}
\label{}
F^{-1}(T) = \frac{1}{\beta}\sinh(\beta T),
\end{equation}
\noindent what means that function $F(\tilde{T})$, using  again (\ref{ct1}), is
\begin{equation}
\label{}
F(\tilde{T}) = \frac{1}{\beta}\mathrm{arcsinh}(\beta \tilde{T}).
\end{equation}
\noindent The full generating function (\ref{can_trans})
\begin{equation}
\label{}
G(\tilde{T}, P) = -PF(\tilde{T}) = -\frac{P}{\beta}\mathrm{arcsinh}(\beta \tilde{T}),
\end{equation}
\noindent reduces equation of motion again to the form
\begin{equation}
\label{}
\ddot{\tilde{T}} - \beta^2\tilde{T} = 0.
\end{equation}
\noindent Here we find the same situation, this equation of motion can be delivered from a quadratic Lagrangian
\begin{equation}
\label{quad2L}\mathcal{L}_{quad}(\tilde{T},\dot{\tilde{T}}) =
\frac{1}{2}\dot{\tilde{T}}^2 + \frac{1}{2}\beta^2\tilde{T}^2.
\end{equation}

Let us stress again that this equivalence is a "local" one, based on (\ref{F}). Keeping in mind that motivation for this study
comes from inflation theory and should be applied to its very beginning (quantum origin and very short period of time),
this simplification seems quite meaningful for further consideration.

\section{Equivalent Lagrangians Quantization revisited}
\label{generalization}

As we noticed in the previously considered simplified cases it is possible
to pass from the non-standard (DBI) Lagrangian to the locally equivalent and "canonical" one.
In both cases, we ended up with the quadratic Lagrangian,
with the potential terms having the "wrong" sign, i.e. the form of an inverted harmonic oscillator Lagrangian \cite{barton}.

Its repulsive ("wrong" sign), negative pressure-like and "antigravitational" effect mimics "dark energy" and offers a
playground for formal speculation and consideration of its real, $p$-adic and adelic origin. This will be discussed elsewhere.
On specualtion of "$p$-adic dark energy" see, for instance, \cite{dragovich06}.

At the (real, i.e. non $p$-adic) quantum level, inverted oscillator system has an energy spectrum,
varying from minus to plus infinity \cite{barton}. So, the state with the lowest energy corresponds to
negative infinite energy, $E = -\infty$. The general solution of Schroedinger equation for the inverted oscillator
can be presented as a linear combination of solutions with definite parity \cite{bermudez}
\begin{equation}
\label{PsiReal}
\Psi(\tilde{T}) = C\Psi_{even}(\tilde{T}) + D\Psi_{odd}(\tilde{T}),
\end{equation}
\noindent where $C$ and $D$ are real constants and $\Psi_{even}$ and $\Psi_{odd}$ are expressed in terms of confluent
hyperbolic functions.

By introducing "annihilation" and "creation" operators, as was done for the harmonic oscillator, one ended up with the so-called generalized eigenstates belonging to the complex energy eigenvalues.
As it is known, the energy eigenvalue $E$ can be a complex number for an unstable system
for which the potential energy does not have a stable stationary point. That is the case here
(see \cite{shimbori} and reference therein for the discussion about mathematical formulations of continuous spectrum 
or complex eigenvalues on real space).

One can ask for which (tachyon-like) potentials present in the DBI
Lagrangian it is possible to get quadratic Lagrangian and with
which sign, i.e. formally with an attractive or repulsive force?

Invoking classical canonical transformation, first of all to simplify equation of motion, we deliver an equation which can be obtained
from the various standard and non-standard type Lagrangians. At the classical level, we can choose the most suitable Lagrangian for the system.
However, at the quantum level, this is not so simple. As different Lagrangiangs can lead to different quantum systems.

Regarding the quantum level, the path integral formulation of quantum mechanics (and also other equivalent formulations) depends
on a Lagrangian chosen to describe the classical system. The arbitrariness in this choice of suitable classical Lagrangian
leads to the so-called quantization ambiguity.

Nevertheless, in the next Section we will discuss the quantum mechanical propagator on both real and $p$-adic spaces
(Archemedean and non-Archemedean ones),
constructed from the most suitable Lagrangian which gives the same equation of motion as the original one.
This is related to the idea \cite{kochan} in which functional integral formula for the quantum propagator could be defined
starting from the given classical equation of motion and, hence, would not refer to the form of the chosen Lagrangian.

\section{Real, $p$-Adic and Adelic Consideration}
\label{p-adic}

As noted in \cite{MoellerZwiebach}, it was shown that
$p$-adic string resembles the bosonic string in such way that its ground
state is a tachyon, whose unstable maximum presumably indicates
the presence of a decaying brane, analogous to the unstable
D25-brane of the open bosonic string theory \cite{sen02}.

A possibility to get successful inflation from rolling $p$-adic
tachyons toward the bounded direction was shown in \cite{Cline}.
This is an opposite and very interesting appearance compared to the real case \cite{kofman},
because the tachyon potential is not flat enough to give a
significant period of inflation. Due to nonlocality and the
nonarchimedean character of $p$-adic space, $p$-adic string tachyon
can roll slowly enough to give many $e$-foldings of inflation.
Nevertheless, it is natural to expect that $p$-adic effects should be of some importance
for the initial phase of inflation and its "very first $e$-foldings".
However, all similar considerations of ($p$-adic) tachyons have been undertaken in
a classical manner and quantum dynamics remains to be studied.

Hence, we start our quantum consideration from Lagrangian
(\ref{quad1L}) or (\ref{quad2L}), which are quadratic with respect
to $\tilde{T}$ and $\dot{\tilde{T}}$. We can write down the
corresponding transition amplitudes (propagators) on real spaces
and using real numbers \cite{morette},
\begin{equation}
{\cal K}_{\infty}(\tilde{T}_2, \tau; \tilde{T}_1, 0)=\sqrt{-\frac{1}{2\pi i
\hbar}\frac{\partial^2 S_c}{\partial \tilde{T}_1 \partial
\tilde{T}_2}}e^{i\frac{S_c}{\hbar}},
\end{equation}
\noindent where $S_c$ is a classical action of the system, which is quadratic with respect to initial and final configuration
$\tilde{T}_1$ and $\tilde{T}_2$. Here, $\tau$ denotes elapsed time.
It can be also written in the form \cite{ModPhys97}
\begin{eqnarray}
\label{real-prop}
{\cal K}_{\infty}(\tilde{T}_2, \tau; \tilde{T}_1, 0)&=&
\lambda_{\infty}\left(-\frac{1}{2h}\frac{\partial^2 S_c}{\partial
\tilde{T}_1
\partial \tilde{T}_2}\right)\left|\frac{1}{h}\frac{\partial^2
S_c}{\partial \tilde{T}_1
\partial \tilde{T}_2}\right|_{\infty}^{1/2}
\times\nonumber\\
&&\chi_{\infty}
\bigg(-\frac{1}{h} S_c(\tilde{T}_2, \tau; \tilde{T}_1, 0) \bigg),
\end{eqnarray}
\noindent where an arithmetic $\lambda$-function and additive character $\chi_{\infty}$ are defined as
\begin{equation}
\lambda_{\infty}(b)=e^{-\frac{i\pi}{4}{\textit sgn}(b)}, \quad \chi_{\infty}(a)=e^{-2\pi ia}.
\end{equation}

Let us discuss the transition amplitude in the $p$-adic case. It
can be formally done by changing the number field, from $R=Q_{\infty}$
to $Q_p$. This means that we will deal with $p$-adic numbers and
complex wave functions of $p$-adic argument \cite{vvz}.

The transition amplitude in $p$-adic case ${\cal K}_p$, for an action quadratic in $\tilde{T}_1$ and $\tilde{T}_2$
(we take $h=1$ for simplicity), as it was shown in \cite{ModPhys97} is
\begin{equation}
\label{Kp}
{\cal K}_p(\tilde{T}_2, \tau; \tilde{T}_1, 0) =
\lambda_p \bigg(-{1\over 2}{\partial^2
S_c\over\partial \tilde{T}_2\partial \tilde{T}_1} \bigg)
\Big\arrowvert{\partial^2
S_c\over\partial \tilde{T}_2\partial \tilde{T}_1} \Big\arrowvert^{1/2}_p
\chi_p
\bigg(-S_c(\tilde{T}_2, \tau; \tilde{T}_1, 0) \bigg),
%\frac{\Big\arrowvert{\partial^2
%S_c\over\partial y_2\partial y_1} \Big\arrowvert^{1/2}_p}
%{\lambda_p \bigg({1\over 2}{\partial^2
%S_c\over\partial y_2\partial y_1} \bigg)}
%\chi_p
%\bigg(-S_c(y_2, T; y_1, 0) \bigg),
\end{equation}
\noindent where the $p$-adic additive character $\chi_p$ is defined as \cite{vvz}
\begin{equation}
\label{chip}
\chi_{p}(a)=e^{2\pi i\{a\}_p},
\end{equation}
\noindent $\{a\}_p$ is the fractional part of the $p$-adic number $a$, while $\lambda_p$ is an arithmetic complex-valued function
(here with a $p$-adic variable), with the following basic properties
\begin{equation}
\label{lambdap}
\lambda_{p}(0)=1, \quad \lambda_{p}(a^2b)=\lambda_{p}(b), \quad |\lambda_{p}(b)|_{\infty}=1,
\end{equation}
\begin{equation}
\label{lambdap1}
\lambda_{p}(a)=1, \quad |a|_p = p^{\mbox{\small ord}(a)} = p^{2\gamma}, \quad \gamma \in Z.
\end{equation}

In this way, the transition amplitude ${\cal K}_p$ for the Lagrangians (\ref{quad1L}), (\ref{quad2L})
has the form
\begin{eqnarray}
\label{padic-prop}
{\cal K}_p(\tilde{T}_2, \tau; \tilde{T}_1, 0) &=&
\lambda_p \bigg(\frac{\gamma}{2\mbox{sinh}(\gamma \tau)} \bigg)
\Big\arrowvert\frac{\gamma}{\mbox{sinh}(\gamma \tau)} \Big\arrowvert^{1/2}_p
%\times
\nonumber \\
&&\times \chi_p
\bigg(-\frac{\gamma}{2} \left(\left(\tilde{T}_1^2+\tilde{T}_2^2\right)
\coth (\gamma \tau)- \frac{2 \tilde{T}_1 \tilde{T}_2}{\mbox{sinh}(\gamma \tau)}\right) \bigg),
\end{eqnarray}
\noindent where
%$\mbox{csch}(\gamma \tau) = 1/\mbox{sinh}(\gamma \tau)$, and
$\gamma$ stends either for $\alpha$ regarding
Lagrangian (\ref{quad1L}) or $\beta$ regarding (\ref{quad2L}), respectively.

The necessary condition for the existence of a $p$-adic and an
adelic model is to find a $p$-adic quantum-mechanical ground state
in the form of $\Omega$-function
\begin{equation}
\label{omega} \Omega(|\tilde{T}|_p)=
\left\{\begin{array}{crl}
1, & \textrm{if} & |\tilde{T}|_p\leq 1\\
0, & \textrm{if} & |\tilde{T}|_p> 1,
\end{array}
\right.
\end{equation}
\noindent which is characteristic function on $Z_p$, the set of $p$-adic integers.
A physical interpretation is that a system (particle) remains in its vacuum (ground) state
as long as it is considered to be inside a "box" with a "natural" unity length (for example Planck length $l_{pl}$), etc.
Note that $\Omega$-function is a counterpart of the Gaussian $\exp{(-\pi \tilde{T}^2)}$ in the real case ($\tilde{T}\in R$), since it is invariant with
respect to the Fourier transform \cite{we-vol}.

Having in mind one of the basic properties of ($p$-adic) propagator and 
the corresponding unitary operator of evolution
\begin{equation}
\label{} \hat{U}(\tau)\Psi_p^{vac}(\tilde{T}_2) = \int_{Q_p}{\cal K}_p (\tilde{T}_2, \tau; \tilde{T}_1, 0)\Psi_p^{vac}(\tilde{T}_1)d\tilde{T}_1,
\end{equation}
\noindent we get for the vacuum state $\Psi_p^{vac}(\tilde{T})=\Omega(|\tilde{T}|_p)$
\begin{equation}
\label{6.1} \int_{Z_p}{\cal K}_p (\tilde{T}_2,\tau;\tilde{T}_1,0)d\tilde{T}_1=
\Omega(|\tilde{T}_2|_p),
\end{equation}
\noindent where the interval of integration becomes restricted to $Z_p$ \cite{ModPhys97}.

The necessary conditions for the existence of ground states in the
form of the characteristic $\Omega$-function are ``dictated`` by
the expression (\ref{6.1}), from which the necessary restriction on
the values of the parameters of the theory can be found
\cite{facta14}. For $p\neq 2$ the necessary conditions are
\begin{equation}
\label{wf1}
\Psi_p^{vac}(\tilde{T})=\Omega(|\tilde{T}|_p), \quad \mbox{for}
\left\{\begin{array}{crl}
&|\tau|_p = 1,& \\
&|\tau|_p < 1,& \quad |\gamma^2 \tilde{T}^2 \tau|_p \leq 1.
\end{array}
\right.
\end{equation}

Note that in $p$-adic quantum mechanics there is degeneration of
the vacuum state. There are other possibilities for the ground
states wave function: "modified" $\Omega$-function
$\Omega(p^\nu|\tilde{T}|_p)$ and $p$-adic Dirac delta-function
$\delta(p^\nu-|\tilde{T}|_p)$, $\nu\in Z$ \cite{we-vol}. They will give another
set of conditions for the existence of ground state and
restrictions for the parameters \cite{ModPhys97}. This
consideration will be discussed elsewhere.

For the theory under consideration, the adelic wave function would be of the form
\begin{equation}
\label{a_ground} \Psi_{ad}(\tilde{T}_a)=
\prod_{v}\Psi_{v}(\tilde{T}_{v})=
\Psi_{\infty}(\tilde{T}_{\infty})\prod_{p\in M}\Psi_p(\tilde{T}_p)\prod_{p\notin M}\Omega(|\tilde{T}_p|_p),
\end{equation}
\noindent where $v=(\infty, 2, 3, ...p,...), $ $M$ is a finite set of primes $p$,
while $\tilde{T}_{\infty}\in R$ and $\tilde{T}_p\in Q_p$ defines an adele
$\tilde{T}_a$, i.e. a sequence of the form
\begin{equation}
\label{adele}
\tilde{T_a} =(\tilde{T}_{\infty}, \tilde{T}_2, \tilde{T}_3, ... \tilde{T}_p, ...),
\end{equation}
\noindent where for all but finitely many $p$, $|\tilde{T}|_p\leq 1$.
The corresponding adelic transition amplitude (quantum propagator) as a non--trivial product
of (\ref{real-prop}) and (\ref{padic-prop}) for all $p$ is \cite{we-vol}
\begin{eqnarray}
\label{adel-prop}
{\cal K}_{ad}(\tilde{T}_2, \tau; \tilde{T}_1, 0) &=&
\prod_{v}{\cal K}_{v}(\tilde{T}_2, \tau; \tilde{T}_1, 0)=
\nonumber \\
&&
{\cal K}_{\infty}(\tilde{T}_2, \tau; \tilde{T}_1, 0)\times\prod_{p}{\cal K}_{p}(\tilde{T}_2, \tau; \tilde{T}_1, 0).
\end{eqnarray}

Note that $\Psi_{\infty}(\tilde{T}_{\infty})$ is the corresponding
real (counterpart of the) wave functions of the theory, and
$\Psi_p(\tilde{T}_p)$ are the $p$-adic wave function (again, for all but finitely many $p$,
$\Psi_p(\tilde{T}_p)=\Omega(|\tilde{T}_p|_p)$). In case of a "true" adelic vacuum state
$\Psi_p(\tilde{T}_p)=\Omega(|\tilde{T}_p|_p)$ for all $p$.

\section{Conclusion}
\label{conclusion}

We considered the DBI-type tachyon scalar field theory, inspired
by a common belief that tachyon field can be used in the frame
of cosmology, in particular in the initial phase of inflation. We
discussed the idea to extend a standard approach using $p$-adic
numbers and ultrametric geometry and spaces or, more generally, 
non-Archimedean ones. This, at the very least, bearing in mind that all results of
experimental and observational measurements are always some
rational numbers, justifies interest in this generalization
\cite{ostrowski, vvz}.

We presented the possibility of passing from the non-standard (DBI)
Lagrangian to the standard and very familiar one (i.e. Lagrangian
for inverted harmonic oscillator). In this paper we presented an
original canonical transformation of a particular form, suitable
for transformation of a class of relevant tachyonic
potentials. We calculated the corresponding quantum propagators using the Feynman path
integral approach in two special cases (exponential and $cosh$
potential) and presented their form on real, $p$-adic and adelic
spaces. A set of conditions for the existence of vacuum states in
$p$-adic and adelic cases, which restricts the allowed domain, for time,
tachyonic field (``distance`` in ``one-dimensional`` classical
mechanical limit of field theory) and parameters present in
considered potentials were obtained. For instance, one can speculate that the tachyonic scalar field exists in an adelic
vacuum state as long as $|\tau|_p < 1$ (for instance, $t_{pl}=1$, or some other natural unity of time) and predict
$|\gamma^2 \tilde{T}^2 \tau|_p \leq 1$. It is obvious that (\ref{wf1}) gives a very interesting constraint on values of
tachyon field $T$, elapsed time $\tau$ and constant parameters of the theory $\gamma$ (i.e. $\alpha$ and $\beta$). 
It can be used for calculation and prediction of values of
(tuning parameter) constants $\gamma$, which determine the shape of potentials (\ref{exp-pot}) and (\ref{cosh-pot}).
%($\gamma = \alpha$ or $\gamma = \beta$, respectively). 
It can also be connected to allowed values of position of "classical particle" counterpart and tachyonic field.

The theory of $p$-adic inflation is not yet ''phenomenological'' enough to predict concrete values of
these constants ($\gamma$, $\tilde{T}$, $\tau$, second line of (\ref{wf1})) and corresponding parameters of inflation. Thoughts
on that, including solution of Friedmann-like equations, will be presented elsewhere \cite{BGDM}.
We find that our approach based on canonical transformation can be applied to inverse power
tachyon potentials \cite{GDM1}. Nearly quadratic systems with a term which could be studied as a
perturbation of a quadratic system is also a very interesting
task.

We find quite interesting that the tachyonic system provides rich
and fruitful cosmological scenarios for inflation, as well as for
dark components, to possess nontrivial quantum dynamics and deserve
further attention and consideration. We would like to underline that 
our approach can be useful in a complete "real" consideration,
when all $p$-adic and adelic extensions are neglected.

Finally, we mention one of the open problems: how to connect wave function of the Universe \cite{we-vol}
with tachyon quantum effects and wave function (\ref{a_ground}), i.e. $\Psi_{\infty}$, $\Psi_{p}$ and $\Psi_{ad}$.

\section{Acknowledgements}

This work was supported by ICTP - SEENET-MTP project PRJ-09
Cosmology and Strings, and by Serbian Ministry for Education,
Science and Technological Development under projects No 176021 and
No 174020. G.S.Dj would like to
thank to CERN-TH, where part of the paper was done, for kind hospitality and support.


\begin{thebibliography}{22}

\bibitem{martin12} J. Martin, V. Vennin and P. Peter, Phys. Rev. {\bf D 86}, 103524 (2012).

\bibitem{martin04} J. Martin and C. Ringeval, Phys. Rev. {\bf D 69}, 083515 (2004).

\bibitem{martin01}
J. Martin and R. Branderberger, Phys. Rev. {\bf D 63}, 123501 (2001).

\bibitem{we-vol}
G.S. Djordjevic, B. Dragovich, Lj.D. Nesic and I.V. Volovich, Int.
J. Mod. Phys. {\bf A 17}, 1413 (2002).

\bibitem{GLjDR}
G.S. Djordjevic, Lj. Nesic and D  Radovancevic, Int.
J. Mod. Phys. {\bf A 29}, 1450155 (2014).

\bibitem{liu}
D.-jun Liu and X.-zhou Li, Phys. Rev. {\bf D 70}, 123504 (2004).
%Cosmological perturbations and noncommutative tachyon inflation

\bibitem{kempf01}
A. Kempf, Phys. Rev. {\bf D 63}, 083514 (2001).

\bibitem{danielsson}
U.H. Danielsson, Phys. Rev. {\bf D 66}, 023511 (2002).

\bibitem{sen02} A. Sen, JHEP {\bf 0204}, 048 (2002).

\bibitem{gibbons}
G.W. Gibbons, Class. Quantum Grav. {\bf 20}, S321 (2003).
%G.W. Gibbons, Phys. Lett. {\bf B 537}, 1 (2002).

\bibitem{picon}
C. Armendariz-Picon, T. Damour and V. Mukhanov, Phys. Lett.
{\bf B 458}, 209 (1999).

\bibitem{garriga}
J. Garriga and V.F. Mukhanov, Phys. Lett. {\bf B 458}, 219 (1999).
%219"1¤7225

\bibitem{kofman}
L. Kofman and A. Linde, JHEP {\bf 07}, 004 (2002).
%Problems with tachyon inflation

\bibitem{Cline}
N. Barnaby, T. Biswas and J. M. Cline, JHEP {\bf 0704}, 056 (2007).

\bibitem{MoellerZwiebach}
N. Moeller and B. Zwiebach, JHEP {\bf 0210}, 034 (2002).

\bibitem{ostrowski}
A. Ostrowski, Acta Math. {\bf 41}, 271 (1918).

\bibitem{vvz}
V.S. Vladimirov, I.V. Volovich and E.I. Zelenov, \textit {p-Adic Analysis
and Mathematical Physics}, World Scientific, Singapore, 1994.

\bibitem{dragovich09}
B. Dragovich, A. Yu. Khrennikov, S.V. Kozyrev and I.V. Volovich,
p-Adic Numbers, Ultrametric Analysis, and Applications, {\bf 1}, No. 1, 1 (2009).

\bibitem{idaqp}
G.S. Djordjevic, B. Dragovich and Lj. Nesic, Inf. Dim. Analys. Quan. Probab. and Rel. Topics {\bf 6}, 176 (2003).

\bibitem{frolov}
A. Frolov, L. Kofman, A. Starobinsky, Phys. Lett. {\bf B 545}, 8 (2002).
%Prospects and problems of tachyon matter cosmology

\bibitem{copeland}
E.J. Copeland, M.R. Garousi, M. Sami and S. Tsujikawa, Phys.
Rev. {\bf D 71}, 043003 (2005).

\bibitem{steer} D.A. Steer and F. Vernizzi, Phys. Rev. {\bf D 70}, 043527
(2004).
%Tachyon inflation: Tests and comparison with single scalar field inflation

\bibitem{fdp}
D.D. Dimitrijevic, G.S. Djordjevic and Lj. Nesic, Fortschr. Phys.
{\bf 56}, No. 4-5, 412 (2008).
%412--417

\bibitem{ddmm} D.D. Dimitrijevic and M. Milosevic, AIP Conf.
Proc. {\bf 1472}, 41 (2012).
%About non standard Lagrangians in cosmology
%doi: 10.1063/1.4748066

\bibitem{kar}
D. Jain, A. Das and S. Kar, AJP {\bf 75}, 259 (2007).
%Path integrals and wavepacket evolution for damped mechanical systems

\bibitem{dvali}
G. Dvali, G.F. Giudice, C. Gomez and A. Kehagias, JHEP {\bf 1108}, 108 (2011).

\bibitem{tetradis}
J. Rizos and N. Tetradis, JHEP {\bf 1204}, 110 (2012).

\bibitem{aip14}
D.D. Dimitrijevic, G.S. Djordjevic, M. Milosevic and D. Vulcanov,
AIP Conf. Proc. {\bf 1634}, 9 (2014).
%On Classical and Quantum Dynamics of Tachyon-like Fields and their Cosmological Implications,

\bibitem{facta14}
D.D. Dimitrijevic, G.S. Djordjevic, M. Milosevic and Lj. Nesic
Facta Universitatis Series: Physics, Chemistry and Technology, Vol {\bf 12}, No 2 (2014).
%DBI-type Tachyons for Inverse {it cosh} Potential

\bibitem{garousi}
M.R. Garousi, Nucl. Phys. {\bf B 584}, 284 (2000).

%\bibitem{franche} P. Franche, R. Gwyn, B. Underwood and A. Wissanji, Phys. Rev. {\bf D 81, 123526 (2010).
%Attractive Lagrangians for Noncanonical Inflation

%\bibitem{gwyn}
%R. Gwyn, M. Rummel and A. Westphal, JCAP {\bf 1312}, 010 (2013).
%Relations between canonical and non-canonical inflation

\bibitem{bilic}
O.F. Piattella, J.C. Fabris, N. Bilic, Class. Quantum Grav. {\bf 31}, 055006 (2014).

%\bibitem{born}
%M. Born, W. Heisenberg and P. Jordan, Ztschr. f. Phys. {\bf 35}, S. 557 (1926); W.
%Heisenberg, Math. Ann. {\bf 95}, 683 (1926); P.~A.~M. Dirac, Proc. Roy. Soc. {\bf A 110}, 561
%(1926).

%\bibitem{anderson}
%A. Anderson and S. Anderson, Annals of Physics, {\bf 199}, 155 (1990);
%A. Anderson, Annals of Physics {\bf 232}, 292 (1994).

\bibitem{musielak}
Z.E. Musielak, J. Phys. A: Math. Theor. {\bf 41}, 055205 (2008).
%Standard and non-standard Lagrangians for dissipative dynamical systems with
%variable coefficients

\bibitem{garousi05}
M.R. Garousi, M. Sami and S. Tsujikawa, Phys. Rev {\bf D 71}, 083005 (2005).
%Constraints on Dirac-Born-Infeld type dark energy models from varying alpha

\bibitem{barton}
G. Barton, Annals of Physics {\bf 166}, 322 (1986).

\bibitem{dragovich06}
B. Dragovich, AIP Conf. Proc. {\bf 826}, 25 (2006).
%$p$-Adic and Adelic Cosmology: $p$-Adic Origin of Dark Energy and Dark Matter

\bibitem{bermudez}
D. Bermudez and D.J. Fernandez, Annals of Physics {\bf 333}, 290 (2013).
%Factorization method and new potentials from the inverted oscillator
%290"1¤7306

\bibitem{shimbori}
T. Shimbori, Physics Letters {\bf A 273}, 1"1¤7-2, 37 (2000);
%Operator methods of the parabolic potential barrier, Physics Letters A, Volume 273, Issues 1"1¤72, 14 August 2000, Pages 37-41,
T. Shimbori and T. Kobayashi, Nuovo Cim. {\bf B 115}, 325 (2000).
%Complex Eigenvalues of the Parabolic Potential Barrier and Gel'fand Triplet
%325-342

\bibitem{kochan}
D. Kochan, Phys. Rev. {\bf A 81}, 022112 (2010).

\bibitem{morette}
C. Morette, Phys. Rev. {\bf 81}, 848 (1951).

\bibitem{ModPhys97}
G.S. Djordjevic and B. Dragovich, Mod. Phys. Lett. {\bf A 12}, 1455 (1997).
%1455"1¤71463

\bibitem{BGDM}
N. Bilic, D.D. Dimitrijevic, G.S. Djordjevic and M. Milosevic,
{\it Inflation parameters for tachyonic power potential} ({\it
work in progress}).

\bibitem{GDM1}
D.D. Dimitrijevic, G.S. Djordjevic and M. Milosevic {\it Classical
Quantum dynamics of tachyonic DBI models with power potential}
(\it{work in progress}).


\end{thebibliography}
\end{document}